\documentclass[%
 reprint,
superscriptaddress,
 amsmath,amssymb,
 aps,
]{revtex4-2}

\usepackage[inkscapelatex=false]{svg}
\usepackage{graphicx}
\usepackage{dcolumn}
\usepackage{bm}
\usepackage{amsmath}
\usepackage{braket}

\begin{document}

\preprint{APS/123-QED}

\title{Hardware Efficient Quantum Kernels Using Multimode Bulk Acoustic Resonators}

\author{Collin C. D. Frink}
\affiliation{Department of Physics, University of Wisconsin– Madison, Madison, WI 53706, USA}


\author{Chaoyang Ti}
\affiliation{Center for Nanoscale Materials, Argonne National Laboratory, Lemont, Illinois 60439, USA}

\author{Stephen K. Gray}
\email{gray@anl.gov}
\affiliation{Center for Nanoscale Materials, Argonne National Laboratory, Lemont, Illinois 60439, USA}
%

\author{Xu Han}
\email{xu.han@anl.gov}
\affiliation{Center for Nanoscale Materials, Argonne National Laboratory, Lemont, Illinois 60439, USA}
\affiliation{Pritzker School of Molecular Engineering, University of Chicago, Chicago, Illinois 60637, USA}

\author{Matthew Otten}
\email{mjotten@wisc.edu}
\affiliation{Department of Physics, University of Wisconsin– Madison, Madison, WI 53706, USA}
\affiliation{Department of Chemistry, University of Wisconsin– Madison, Madison, WI 53706, USA}
\email{mjotten@wisc.edu}


\date{\today}

\begin{abstract}
The kernel trick is a widely applicable technique in machine learning domains that maps datasets that are difficult to classify into a computationally friendly feature space. As the dimension of the dataset scales, these kernel calculations can quickly become computationally intractable or data inefficient. In this work, we extend prior efforts in quantum kernel design for Kerr nonlinear devices by implementing time-dependent simulations of a Kerr-qubit coupled to acoustic resonators. For experimentally feasible parameters, we demonstrate that the Kerr nonlinearity directly induces non-classical behavior in the multimode system, which we use to define and analyze a quantum-enhanced kernel. Finally, we present a brief scaling characterization that demonstrates the computational intractability of classically simulating the kernel as the number of resonators scales.

\end{abstract}

\maketitle


\section{\label{sec:Introduction}Introduction}

Machine learning (ML) has been one of the most dominant research fields in the past decade, with applications including healthcare, cybersecurity, agriculture, and many more \cite{Sarker2021}. Real-world datasets contain highly complex correlation structures and mappings from an input feature space $X$ to an output label space $L$ that are usually further obscured by noise on both sides \cite{Song2020LearningFN}. One technique used in several ML models to combat this is the kernel trick, which leverages a positive definite kernel function $K:X\times X \to \mathbb{R}$ that implicitly maps $X$ into a feature space that is more favorable for label classification \cite{Hofmann2008}. In many non-trivial kernel-based methods, however, as the application size scales, the computing costs quickly increases while the data efficiency inversely decreases \cite{tradeoff}.

Superposition is a fundamental feature of quantum computing, which allows $n$ quantum modes of dimension $d$ to represent $O(d^n)$ computational states simultaneously. Given this advantage, it is natural to try fully quantum or hybrid classical-quantum techniques to help mitigate the scaling issues purely classical ML methods face. In fact, quantum-enhanced kernel methods have gained increasing attention in the last decade, with extensive analysis of both classically-inspired and novel quantum kernel matrix schemes \cite{Schnabel2025,jiang2025benchmarking,schuldQuantumMachineLearning2019,wangComprehensiveReviewQuantum2024}. In this work, we extend \textit{quantum Kerr learning}: a  quantum kernel approach that embeds input data samples into the Hamiltonian of a nonlinear Kerr qubit \cite{Liu2023}. 

\section{\label{sec:Device} Methods}

\begin{figure*}
\includegraphics[width=\textwidth]{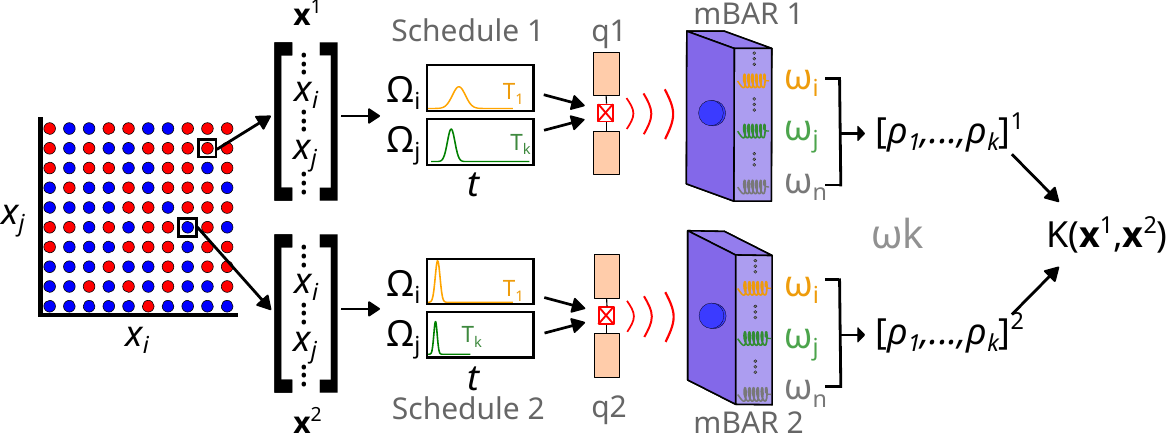}
\caption{\label{fig:device} \textit{Schematic of the end-to-end  quantum-enhanced kernel calculation pipeline.} Two samples $(\textbf{x}^1, \textbf{x}^2)$ from a dataset with class labels $\{red, blue\}$ and features $(x_i, x_j)$ are mapped to pulse schedules. Each sample can be encoded into the schedule $\textbf{$\bf \Omega$}$ and list of measurement times $\textbf{T}$. The drives are applied to a Kerr nonlinear qubit $q$ and selectively excite resonator modes ($\omega$) in multimode bulk acoustic resonators (mBAR) via piezoelectric transduction. 
Certain reduced density matrices,
$\rho$, of the mBARs are  inferred from the quantum dynamics
and then used as an input to the kernel calculation $K\bf{(x^1}, \bf{x^2})$.} 
\end{figure*}

\subsection{Hardware}

We consider a multimode quantum system with Kerr nonlinearity, comprising of multiple mechanical modes that are strongly coupled to a qubit (as shown in Fig. \ref{fig:device}). Experimentally, this architecture can be realized using a multimode bulk acoustic resonator (mBAR) integrated with a superconducting qubit \cite{Chu2017, Chu2018}. The mBAR supports a series of gigahertz-frequency phonon modes separated by megahertz free spectral range (FSR), providing a versatile multimode platform for engineering complex quantum states. When interfaced with a high-coherence qubit, the system acquires a Kerr nonlinearity (to the first order approximation) that mediates controllable nonlinear interactions among the phonon modes.

The coupling between the qubit and the mBAR can be established via the piezoelectric effect \cite{Han2016, Han2022}: the qubit’s electric field induces strain in the resonator, exciting acoustic modes; and vice versa, the mechanical deformation generates charges and modulates the qubit’s electric field. This mechanism realizes a Jaynes–Cummings–type interaction between the qubit and individual phonon modes. The qubit’s Kerr nonlinearity further enables tunable nonlinear interactions across the multimode phonon manifold, offering opportunities for creating entanglement and complex multimode quantum states beyond what is possible with purely linear couplings. Without Kerr nonlinearity, the qubit reduces to a harmonic oscillator while the system Hamiltonian reduces to a linear form: in the rotating frame, the mechanical modes become mutually decoupled, and the qubit interacts linearly with each mode. When the Kerr nonlinearity is activated, we compare the effective Kerr coefficient under different drive amplitudes and drive frequencies.

In experiment, quantum control and measurement can be performed using a circuit quantum electrodynamics (cQED) architecture. In particular, the overlap between two states can be measured by two Hadamard gates (H) with a controlled-SWAP (CSWAP) gate inserted in the middle. Then the overlap can be inferred by measuring the probability of the qubit's final state in $\ket{g}$ or $\ket{e}$ \cite{Liu2023}.

\subsection{Device Hamiltonian}

The time-dependent Schr\"odinger equation defines the evolution of a quantum state $|\psi(t)\rangle$ when acted upon by a Hamiltonian operator $\hat H(t)$ \cite{sakurai_modern_2010}

\begin{equation}
\label{eqn:tdschro}
    i\hbar \frac{d}{dt}|\psi(t)\rangle = \hat H(t)|\psi(t)\rangle.
\end{equation}

\noindent Under the rotating-wave approximation (RWA), the Hamiltonian $\hat H$ of our multimode quantum system can be written  as

\begin{equation}
    \label{eqn:dev}
    \hat H = \hat H_{\mathrm{lin}} + \hat H_{\mathrm{kerr}} + \hat H_{\mathrm{drive}},
\end{equation}

\noindent where each term is expanded as a sum of bosonic operators

\begin{subequations}
\label{eqn:dev_exp}
\begin{gather}
\hat H_{\text{lin}} 
  = -\Delta_q \hat a^\dagger \hat a
    + \sum_{i=1}^n \left[
      -\Delta_i \hat b_i^\dagger \hat b_i
      + g_i(\hat a^\dagger \hat b_i + \hat b_i^\dagger \hat a)
    \right],
  \label{eqn:dev_expa}
\\[4pt]
\hat H_{\text{kerr}}
  = -K_{\text{err}} \hat a^{\dagger 2}\hat a^{2},
  \label{eqn:dev_expb}
\\[4pt]
\hat H_{\text{drive}}
  = \sum_{j=1}^m \Omega_j(t)\bigl(
      \hat a^\dagger e^{i \delta_j t}
      + \hat a e^{-i \delta_j t}
    \bigr).
  \label{eqn:dev_expc}
\end{gather}
\end{subequations}
$\hat a$ and $\hat b_i$ are the annihilation operators of the qubit and the $i^{th}$ resonator mode, respectively. We consider $n$ resonator modes detuned by $\Delta_i$ from the frame of reference chosen in the RWA. The qubit is detuned from the frame by $\Delta_q$ and is defined by a Kerr nonlinearity of strength $K_{err}$. We apply $m$ pulses of drive strength $\Omega_j(t)$ and frequency $\delta_j$ to the qubit, which couples to the resonator modes via a piezoelectric coupling strength of $g_i$. Although a device can be engineered with a range of mode detunings, $K_{err}$ strengths, and coupling strengths, for all simulations considered here we fix the device parameters and encode our data samples $\boldsymbol{x}^i$ into the pulse schedules and measurement times $\textbf{T}^i$=$[T_1,...,T_k]$ as illustrated in Fig. \ref{fig:device}. We note that the index $k$ is distinguished from $m, n$ as multiple wavefunctions per encoding can be fed into the kernel calculation $K(\textbf{x}^1, \textbf{x}^2)$ by measuring at different times.

For the device and drive parameter ranges we consider, the linear phases in the wavefunctions of the full system propagate much faster than the effective time of the Kerr shear. We negate this by aligning the drive times and rotating into the interaction picture with respect to $H_{\mathrm{lin}}$ \cite{sakurai_modern_2010}. After this rotation, for each sample $\boldsymbol{x}^i$ we 
obtain $k$ 
reduced density matrices
$\boldsymbol{\rho}^i=[\rho^i_1,...,\rho^i_k]$ that can be used as inputs into our kernel calculations.

All numerical simulations of Eq. (\ref{eqn:dev}) are performed in Julia \cite{bezanson2012} using the QuantumToolbox software package \cite{QuantumToolbox.jl2025}. The qubit is numerically truncated to four dimensions, while the resonator modes are truncated to nine dimensions. All Gaussian pulses with a center $t_c$ and standard deviation $\sigma$ are truncated to $\Omega(t)=0$ for times $|t - t_c|\geq3\sigma$. Numerical convergence is verified by increasing the number of truncation dimensions, decreasing the timestep size $\Delta t$, and increasing the included range of each of the pulses.

\subsection{Kernel Calculation}

\label{section:kernel1}

Given an encoding of data samples $(\textbf{x}^1$, $\textbf{x}^2)$ into respective pulse schedules $(\textbf{$\bf \Omega$}^1$, $\textbf{$\bf \Omega$}^2)$ and measurement times $(\textbf{T}^1$, $\textbf{T}^2)$ we most generally define the Kernel matrix element $K_{12}$ as 

\begin{equation}
    \label{eqn:k12}
    K_{12} = K(\boldsymbol{\rho}^1(\textbf{T}^1, \boldsymbol{\Omega}^1), \boldsymbol{\rho}^2(\textbf{T}^2, \boldsymbol{\Omega}^2)).
\end{equation}

Although $K_{12}$ can be any function of $(\boldsymbol{\rho}^1, \boldsymbol{\rho}^2)$, we simplify the expression by encoding each feature $x_i$ into a singular drive pulse amplitude $\Omega_i(t)$ with a drive frequency on resonance with the $i^{th}$ resonator mode at $\omega_i$. That is, for a data sample of dimension $d$, we have $d=m = n=k$. This facilitates the following kernel definition

\begin{equation}
\label{eqn:k12v2}
    K_{ij} = \prod_{l=1}^d F(\rho^1_l, \rho^2_l),
\end{equation}

\noindent where $F$ is the Uhlmann Fidelity \cite{Nielsen_Chuang_2010} defined as

\begin{equation}
    \label{eqn:fidel}
    F(\rho, \sigma) = \text{Tr}[\sqrt{\sqrt{\rho} \sigma \sqrt{\rho}}~ ].
\end{equation}
Note that the fidelity reduces to the magnitude of the overlap between the two pure states without entanglement. In general, we cannot write Eq.\,(\ref{eqn:fidel}) as an overlap because tracing out the qubit may entangle the mBAR modes.

\subsection{Support Vector Machines}

There are several ML models that can be enhanced via the kernel trick, including spectral clustering, Gaussian processes, and support vector machines (SVM)--our model of choice in this manuscript \cite{Hofmann2008}. In the feature space implicitly mapped to by $K$, support vector machines are linear classifiers that maximize margin between two label classes $L_1, L_2$ \cite{Cervantes2020}. A full SVM formulation is available in Ref. \cite{Cervantes2020}.

\section{\label{sec:Results} Results}

\begin{figure}[b]
\includegraphics[width=\columnwidth]{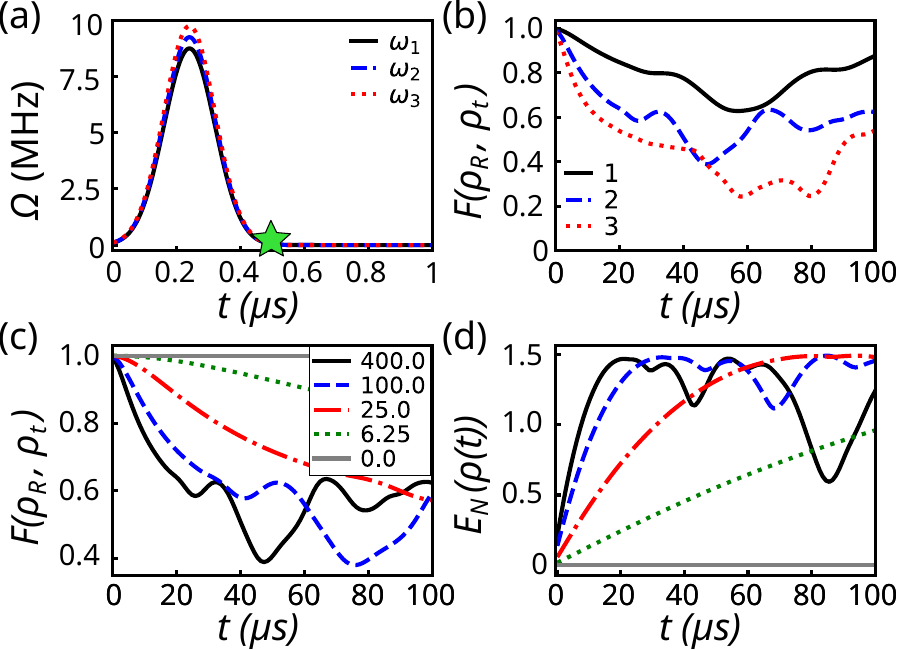}
\caption{\label{fig:entanglement} \textit{Demonstration of entanglement for $K_{err} > 0.$} (a) Gaussian pulses $\Omega_i(t)$ applied to the $K_{err}$ qubit that selectively excite resonators $\omega_i$ for $n\in\{1, 2, 3\}$ with respective line-styles $\{$solid black, dashed blue, dotted red$\}$. Green star is placed at reference time $t_R$ which indicates the end of the pulses.  (b) Fidelity $F(\rho_R, \rho_t)$ of the mBAR state $\rho_t$ at time $t$ with respect to $\rho_R$ at $t_R$ for the number of resonators included $n\in\{1, 2, 3\}$. The device parameters are: $K_{err}/2\pi=400$ MHz, $\Delta_q/2\pi = 100$ MHz, $\Delta_i/2\pi=[10, -10, -30]$ MHz, $g/2\pi=8$ MHz.  As more resonators $\omega_i$ are added to the mBAR and excited by $\Omega_i$, the effective $K_{err}$ of the system increases. (c) Fidelity $F(\rho_R, \rho_t)$ of the mBAR state $\rho_t$ with respect to $\rho_R$ for $n=2$. The Kerr strength of the qubit is varied: $K_{err}/2\pi\in[400., 100., 25., 6.25, 0.]$ MHz with respective line-styles $\{$solid black, dashed blue, dot-dashed red, dotted green, solid grey$\}$. (d) Log-negativity $E_N(\rho_t)$ over $t$ for different $K_{err}$ strengths. For all $K_{err}\neq0$, positive $E_N$ indicates entanglement of the two resonators.}
\end{figure}

\subsection{Kerr-induced Quantum Entanglement}

Figure \ref{fig:entanglement} provides a demonstration of Kerr-induced entanglement in the device. The device parameters are $\Delta_q/2\pi=100$ MHz, $g_i/2\pi=8$ MHz, $\Delta_1/2\pi = 10$ MHz, $\Delta_2/2\pi = -10$ MHz, and $\Delta_3/2\pi = -30$ MHz. First, Fig. \ref{fig:entanglement}(a) depicts a drive schedule consisting of three Gaussian pulses $\Omega_i(t) $ resonant with the three acoustic modes $\omega_i$, where the drive amplitudes are

\begin{equation}
\label{eqn:gaussian}
    \Omega_i(t) = \Omega_i~ \exp[{\frac{(t - t_{p})^2}{2\sigma^2}}].
\end{equation}

\noindent Here, $\sigma=0.08 ~\mu s$, $t_p=3\sigma$, $\Omega_1/2\pi=8.75$ MHz, $\Omega_2/2\pi=9.25$ MHz, and $\Omega_3/2\pi=9.75$ MHz. Although we bump up the drive amplitude for each $\omega_i$ to account for the increasing relative detuning between the qubit and $\omega_i$, we note that the population of the acoustic modes is largest in $i=1$ and $\langle\hat n_{\omega_1} \rangle > \langle\hat n_{\omega_2} \rangle > \langle\hat n_{\omega_3} \rangle$. The FSR of the acoustic modes is chosen to be 20 MHz to suppress population exchange as the modes oscillate. Finally, we define the reference state of the mBAR $\rho_R$ as the state immediately after the drives, denoted by the green star in Fig. \ref{fig:entanglement}(a).

Figure\,\ref{fig:entanglement}(b) examines the trajectories of $\rho(t)=\rho_t$ over time with respect to $\rho_R$ for a device with a Kerr-qubit of strength $K_{err}/2\pi=400$ MHz and $n\in\{1~ (\text{solid black}), 2~ (\text{dashed blue}),3~ (\text{solid black})\}$ acoustic modes. Despite populating each additional acoustic mode less than the previous, we still see a faster and larger divergence in $F(\rho_R, \rho_t)$ as $n$ increases. Interestingly, we nearly see a full phase recovery to $F(\rho_t, \rho_R)=1$ in the $n=1$ case.

Figure\,\ref{fig:entanglement}(c) also examines the trajectories of $\rho_t$ over time with respect to $\rho_R$, but now fixes $n=2$ and varies $K_{err}$. For $K_{err}=0$ MHz (solid grey), in the interaction picture of $\hat H_{\mathrm{lin}}$, by definition states $\rho_R \equiv \rho_t$, so $F(\rho_R,\rho_t)=1$. Increasing $K_{err}/2\pi\in\{6.25, 25,100,400\} $ MHz (with respective line-styles $\{$dotted green, dot-dashed red, dashed blue, solid black$\}$) numerically demonstrates a larger effective $K_{err}$, but interestingly does not change the trajectory of $F(\rho_R,\rho_t)$, only speeds it up.

Figure\,\ref{fig:entanglement}(d) takes the trajectories of $\rho_t$ from Fig. \ref{fig:entanglement}(c) and quantifies the entanglement of the two acoustic modes via the logarithmic negativity,

\begin{equation}
    \label{eqn:logneg}
    E_N(\rho_{w_1, \omega_2}) = \log_2(||\rho^{T_{\omega_1}}||_1) = \log_2(\sum_{\lambda_i < 0}\lambda_i).
\end{equation}

\noindent $E_N(\rho_t) > 0$ indicates entanglement between the two acoustic modes \cite{vidal_werner_2002}. Examining Fig. \ref{fig:entanglement}(d), again for $K_{err}=0$ we have $\rho_R \equiv\rho_t$. Since $\rho_R$ is simply an outer-product of two coherent states both excited in a short time, there is no entanglement between $\omega_1$ and $\omega_2$. As $K$ is increased, however, we immediately see entanglement throughout the trajectories of $\rho_t$. This indicates that a quantum kernel defined by the wavefunctions $\rho_t$ of our device will express non-classical behavior.

\subsection{Quantum Enhanced Kernel}

Figure \ref{fig:qml} defines and analyzes the performance of our quantum kernels on favorable synthetic 2-dimension datasets. First we define an encoding of a data sample $\boldsymbol{x}^i$ into a schedule of pulses $\boldsymbol{\Omega}^i$ and measurement times $\boldsymbol{T}^i$. We then define a procedure to generate favorable synthetic datasets because classical kernels, like the radial basis kernel (RBF), exhibit near-perfect performance on realistic low-feature datasets. Finally, we analyze the performance of our quantum kernels with respect to classical benchmarks. 

\subsubsection{Dataset Encoding}

As indicated in Section \ref{section:kernel1}, we assign each feature $x^i_j$ to a drive amplitude $\Omega^i_j$ and measurement time $T^i_j$ where $i,j\in\{1, 2\}$. Note that at each measurement time $T^i_j$ we measure both resonators $\omega_1,$ $\omega_2$ to include cross-mode Kerr effects in our kernel. 

We encode each data feature $x^i_j \in[0, 1]$ linearly into its respective pulse amplitude $\Omega_j^i(t)$ \textit{and} measurement time $T^i_j$ with

\begin{subequations}
\label{eqn:linearencode}
\begin{gather}
  \Omega_j^i = \Omega_{\min} + (\Omega_{\max} - \Omega_{\min}) x_j^i, \\
  T_j^i      = T_{\min} + (T_{\max} - T_{\min}) x_j^i .
\end{gather}
\end{subequations}

\noindent Each pulse is a Gaussian defined by $\sigma=80 ~\mathrm{ns}$ and $t_p=3\sigma$. We set $\Omega_{min}/2\pi=7.5$ MHz, $\Omega_{max}/2\pi=7.6$ MHz, $T_{min} = (5 + 3\sigma) ~\mathrm{\mu s}$ and $T_{max} = (95 + 3\sigma)~\mathrm{\mu s}$. The small population dependence and long device oscillation time ensure an effective Kerr shear in our device. 

We also tweak the device parameters to increase the effective Kerr strength and use $\Delta_q/2\pi = 70$ MHz, $\Delta_1/2\pi=10$ MHz, $\Delta_2=-15$ MHz, and $g/2\pi=10$ MHz. The strength of the Kerr-nonlinearity $K_{err}$ is analyzed for $0 \leq K_{err}/2\pi \leq 400$ MHz.

\subsubsection{Dataset generation}

To generate our synthetic datasets, we first define a global mesh $X_m\subset[0,1]\times[0,1]$ of 75$\times$75 equally-spaced points. We randomly select two reference points in the mesh, assign each a unique label $l\in\{1, 2\}$ and compute $\boldsymbol{\rho}^{ref_1}$ and $\boldsymbol{\rho}^{ref_2}$ as depicted in Fig. \ref{fig:device}. For every sample $\boldsymbol{x}^i\neq \boldsymbol{x}^{ref_l}$, we compute $K_1(\boldsymbol{\rho}^i, \boldsymbol{\rho}^{ref_1})$, $K_2(\boldsymbol{\rho}^i, \boldsymbol{\rho}^{ref_2})$ and assign $\boldsymbol{x}^i$ a class label $l\in\{1, 2\}$ according to $K_1>K_2\implies l=1$ and vice versa. Training sets of size $2^s$ are randomly selected from $X_m$.

\subsubsection{QSVM Performance}
\begin{figure}[h]
\includegraphics[width=\columnwidth]{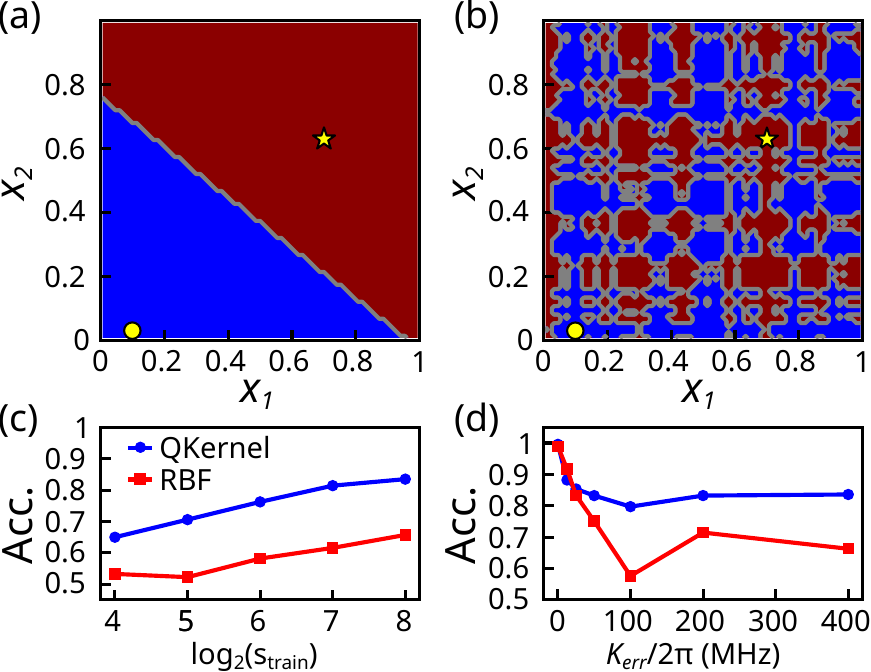}
\caption{\label{fig:qml} \textit{Performance of quantum kernels on favorable synthetic 2D datasets.} (a) Synthetic dataset generated from a device with $K_{err}=0$, $\Delta_q/2\pi = 70$ MHz, $g/2\pi=10$ MHz, and two resonators with detunings $\Delta_1/2\pi=10$ MHz and $\Delta_2/2\pi=-15$ MHz. Reference points for the red, blue classes are indicated by the yellow star, circle (respectively). (b) Same as (a) but $K_{err}/2\pi = 400$ MHz. (c) Performance of the quantum-enhanced and RBF kernels on the  dataset from (b) as a function of the training size $s_{train}$. (d) Evaluation of the quantum-enhanced and RBF kernel performance on datasets generated from device in (a) with varied $K_{err}$. Training set size is fixed at 256 samples. As $K_{err}$ increases, the dataset transforms from (a) to (b), resulting in an increasingly worse performance of the classical RBF kernel compared to its quantum counterparts.}

\end{figure}
Randomly selecting the reference points $\{\boldsymbol{x}^{ref1}=(0.1, 0.04285), \boldsymbol{x}^{ref2}=(0.7, 0.64285)\}$ labeled by $\{$yellow circle, yellow star$\}$ gives the classes $\{$blue, red$\}$ respectively. Figure \ref{fig:qml}(a) and \ref{fig:qml}(b) depict the synthetic datasets generated for $K_{err}/2\pi=0$ and 400 MHz, respectively. Because the only difference in the simulations is the $K_{err}$ strength and Fig. \ref{fig:entanglement} demonstrated entanglement in a device with weaker effective $K_{err}$, we argue that source of the non-trivial boundaries in Fig. \ref{fig:entanglement}(b) is non-classical in nature.

Figure \ref{fig:qml}(c) evaluates the performance of a quantum kernel (blue circle) and a classical RBF kernel (red square) on the dataset in Fig. \ref{fig:qml}(b) as a function of the training-set size $s_{train}$. Data encoding and device parameters for the quantum kernel (Eq. \eqref{eqn:k12v2}) are fixed as defined for Fig. \ref{fig:qml}(b). For each training-set size, three sets are randomly selected from the global mesh and their respective prediction accuracies over the entire global mesh are averaged. The C hyper-parameter for every SVM evaluated is tuned on a log-spaced grid of 50 points with the bounds $-2\leq \log_{10}(C) \leq1$. Additionally, the $\gamma$-parameter is found to be optimized within a linear-spaced grid of 50 points over the bounds $1\leq\gamma\leq1000$. 

There is a clear out-performance of the quantum kernel over the RBF kernel in Fig. \ref{fig:qml}(c) across all training-set sizes. In fact, the RBF kernel needs 256 training points to match the performance of the quantum kernel on 16 training points. Again, this indicates $K_{err}$-induced non-classical structure within the quantum kernel that enables a more data-efficient expressivity of the synthetic dataset compared to that of the RBF kernel.

Finally, we fix the training-set size to 256 samples and vary $0\leq K_{err}/2\pi\leq400$ MHz. Fig. \ref{fig:qml}(d) shows the performance the quantum kernel on its respective synthetic dataset (as in Fig. \ref{fig:qml}(c)) compared to the RBF kernel. Each data-point for all kernels averages the results from two training sets randomly selected from the global mesh. All trials are hyperparameter tuned as in Fig. \ref{fig:qml} with the exception that for quantum kernels $K_{err}=0$, we include an additional linear grid of 500 points for $1\leq C \leq3\times10^6$. 

For all $K_{err}/2\pi>12.5$ MHz in Fig. \ref{fig:qml}(d), the RBF kernel again is out-performed by its quantum counterpart given by Eq.  ($\ref{eqn:k12v2}$). In the limit such that $K_{err}=0$, there is no entanglement between the outer modes themselves or with the qubit. This implies Eq. (\ref{eqn:k12v2}) reduces to the squared-overlap of coherent states defined only by the drive-amplitudes $\boldsymbol{x}^i$ is encoded into. Assuming proper tuning of the C-hyperparameter, this reduces to the RBF kernel \cite{otten2020quantummachinelearningusing}. Figure \ref{fig:qml}(d) not only demonstrates this at $K_{err}=0$, but also shows that phase shears from $K_{err}>0$ directly induces non-classical, \textit{non-Gaussian} structure into the quantum kernels. 

\subsection{Scaling}

Table \ref{tbl:scale} presents an overview of the simulation complexity trends for two-level qubit-only hardware and numerical results of single-threaded classical simulations of one $K_{ij}$, the $(i,j)^{th}$ entry of the kernel matrix $K$. For $n_q$, the number of qubits necessary to simulate our device, we simply evaluate the linearly-scaling space-complexity case (\ref{eqn:qsim}),

\begin{equation}
\label{eqn:qsim}
    n_q = \lceil \log_2(2(n_{dim})^{n})\rceil = \lceil 1 + n \log_2(n_{dim}) \rceil~~~,
\end{equation}

\noindent and $t_c$, the classical simulation time to compute a kernel matrix entry $K_{i,j}$ is as defined for Fig. \ref{fig:qml}. We set all measurement times to 50 $\mu s$, and $\Omega_i/2\pi= 5.$ MHz for $\sigma=80~ns$. We see clear exponential scaling of the classical simulation time in $n$ and at $n_{dim}$ for $n=4$.
In the physical mBAR device, the time to obtain the data is expected to scale only quadratically with the 
number of sub-systems, because we will need to estimate the pairwise fidelities between all the modes. This is 
an exponential speedup over the classical simulation times shown here. We note that these are exact simulations; 
approximate simulations will potentially be cheaper, but the demonstrated entanglement structure implies that it will be difficult to simulate even with approximate techniques~\cite{oh2024classical}.

\begin{table}[h]
\caption{\label{tbl:scale}\textit{ Simulation complexity as $n$, the number of resonators in the mBAR, scales.} $n$ and $n_{dim}$, the truncation dimension of the resonators, are used to calculate the number 2-level qubits needed to simulate the device. Two classical, single-threaded calculations $K_{ij}$ for the device featured in Fig. \ref{fig:qml} are also timed $t_c$ and included.}
\begin{ruledtabular}

\begin{tabular}{cccc}

 $n$ & $n_{dim}$ &$n_q$ & $t_c$ (s) \\
\hline
2& 6 & 7 & 4.7 \\
2& 9 & 8 & 9.6 \\
2& 12 & 9 & 12.4 \\
3& 6 & 9 & 14.8 \\
3& 9 & 11 & 40.2 \\
3& 12 & 13 & 63.6 \\
4& 6 & 12 & 160.5 \\
4& 9 & 14 & 773.3 \\
4& 12 & 16 & 2546.4 \\

\end{tabular}
\end{ruledtabular}

\end{table}

\section{Conclusions}

Our manuscript extends a prior work in quantum Kerr learning \cite{Liu2023} by defining, simulating, and analyzing an end-to-end quantum kernel calculation pipeline that is implemented by a Kerr-nonlinear qubit coupled to acoustic resonators. We demonstrate that the qubit directly induces non-classical behavior into the density matrices that represent the acoustic resonators. We then numerically demonstrate an advantage of our quantum kernels over the classical RBF kernel in performance on favorable synthetic datasets.  We propose the following research directions as extensions of this work.

\textit{a.} Experimental Implementation. Ref. \cite{Liu2023} discussed the natural extension of experimental implementation and theoretical claims of complexity. Because of the high mode density and compact device footprint, the mBAR-qubit system offers a promising hardware-efficient platform for high-dimensional quantum machining learning.

\textit{b.} Kernel Matrix Definition. In this work, we restrain ourselves to short, Gaussian pulses that effectively act as coherent shifts independent of whether $K_{err}$ is turned on or not. In general, pulses can be shaped with respect to a desired encoding which can introduce non-trivial linear oscillations and more non-linear $K_{err}$ effects. We also restrict ourselves to relatively simple definitions of $\boldsymbol{\rho}$ and $K(\boldsymbol{\rho}^1, \boldsymbol{\rho}^2)$, leaving room for perhaps more complex, expressive schemes.

\textit{c.} Gradient Descent. In this work, we only consider optimization-free kernel matrix calculations. However, as Ref. \cite{Liu2023} points out, the pulse parameters could be treated as model-parameters to optimize via gradient descent. It remains open how well this technique generalizes to the multimode system, even in the case of a simplified encoding scheme as presented above.

\textit{d.} Noise. In this work, we only consider noiseless wavefunction calculations and ideal fidelity calculations. However, in practice, all modern quantum systems have significant quantum noise that decoheres the quantum state. It is uncertain the effects different noise processes will have on the Kerr shear evolution. Additionally, shot noise in the fidelity calculations introduces another error source that is not fully analyzed in this manuscript.

\textit{e.} Complexity. As discussed in Ref. \cite{Liu2023}, it remains unclear if there are rigorous complexity statements to be made for our optimization-free scheme. It would also be interesting to formally prove any complexity statements distinguishing or unifying qubit-only vs. hybrid qubit-multimode hardware in the asymptotic computational regimes.

\section*{Acknowledgments}
Work performed at the Center for Nanoscale Materials, a U.S. Department of Energy Office of Science User Facility, was supported by the U.S. DOE, Office of Basic Energy Sciences, under Contract No. DE-AC02-06CH11357.

\bibliography{main}

\end{document}